\documentclass[manuscript,screen,sigconf]{acmart}
\AtBeginDocument{%
  \providecommand\BibTeX{{%
    \normalfont B\kern-0.5em{\scshape i\kern-0.25em b}\kern-0.8em\TeX}}}
\usepackage{listings}
\usepackage{subfigure}

\begin{document}

\title[Tale of Seven Alerts]{Tale of Seven Alerts: Enhancing Wireless Emergency Alerts (WEAs) to Reduce Cellular Network Usage During Disasters}

\author{Demetrios Lambropoulos}
\orcid{}
\affiliation{%
  \institution{WINLAB, Rutgers University} 
  \streetaddress{}
  \city{}
  \state{}
  \postcode{}
  \country{}}
\email{dpl60@rutgers.edu}
\author{Mohammad Yousefvand}
\affiliation{%
  \institution{WINLAB, Rutgers University}
  \streetaddress{}
  \city{}
  \state{}
  \country{}}
\email{my342@winlab.rutgers.edu}
\author{Narayan Mandayam}
\affiliation{%
 \institution{WINLAB, Rutgers University}
 \streetaddress{}
 \city{}
 \state{}
 \country{}}
\email{narayan@winlab.rutgers.edu}

\renewcommand{\shortauthors}{Lambropoulos, et al.}

\begin{abstract}
  In weather disasters, first responders access dedicated communication channels different from civilian commercial channels to facilitate rescues. However, rescues in recent disasters have increasingly involved civilian and volunteer forces, requiring civilian channels not to be overloaded with traffic. We explore seven enhancements to the wording of Wireless Emergency Alerts (WEAs) and their effectiveness in getting smartphone users to comply, including reducing frivolous mobile data consumption during critical weather disasters. We conducted a between-subjects survey (N=898), in which participants were either assigned no alert (control) or an alert framed as Basic Information, Altruism, Multimedia, Negative Feedback, Positive Feedback, Reward, or Punishment. We find that Basic Information alerts resulted in the largest reduction of multimedia and video services usage; we also find that Punishment alerts have the lowest absolute compliance. This work has implications for creating more effective WEAs and providing a better understanding of how wording can affect emergency alert compliance.  
\end{abstract}

\settopmatter{printacmref=false}
\setcopyright{none}
\renewcommand\footnotetextcopyrightpermission[1]{}
\pagestyle{plain}

\maketitle

\section{Introduction}
Mobile device ownership has been steadily on the rise, especially as our remedial tasks rapidly integrate into these devices. As of 2019, $96\%$ of U.S. adults currently own any cellphone type, and $81\%$ own a smartphone~\cite{Demograp89:online}. As technology enhances, the number of bandwidth-intensive tasks a smartphone can perform allows smartphone users to consume more network bandwidth in shorter time durations. The traffic expected to be going through U.S. cellular networks in 2021 is around $5.57$ Exabytes per month, with $343.9$ million smartphones~\cite{UnitedSt40:online}.

To handle the increase in cellular traffic and a continuous stream of newly registered devices, the number of cellular towers has been increased by around $349344$ towers in $2018$, a $235\%$ increase from $2000$~\cite{CTIA201956:online}. On average, these towers are handling approximately $2.9$ thousand simultaneous mobile users per tower. Among these simultaneous users, most 4G cellular towers will be unable to handle higher than 230 simultaneous video streams~\cite{asheralieva2014two} and 5G towers, if deployed, up to 4,700 simultaneous videos simultaneous video streams~\cite{TheTrans29:online}. 

Means of communication have the possibility of degrading during severe weather conditions, reducing the cellular towers' capacity. It is critical that during these times of disaster, smartphone users reduce frivolous usage of tasks, reserving the network for civilians making emergency communications. Wireless Emergency Alerts (WEAs) are the current standard from the Federal Communication Commission (FCC) and Federal Emergency Management Agency (FEMA), in alerting mobile users of intermittent weather conditions and the precautionary measures that need to be adhered to. 

The studies' primary objective is to find the most efficient wording for WEAs to reduce non-essential traffic in cellular networks reserving network resources for first responders and civilians to assist with rescue and recovery operations in emergency disasters. With the assistance of a team of professors in our university's Psychology Department, we designed seven alert types that we believe can replace the current WEAs more effectively. The seven alert types we developed are basic information, altruism, multimedia, negative feedback, positive feedback, reward, and punishment.

This paper first gives a brief overview of the recent history of WEAs, how WEAs are sent, and the current state of First Responder networks. We will then go on to review the literature on warning psychology and the appropriate design of alert messages. After reviewing the literature, the design of our survey and our seven alerts are discussed in depth. Finally, we discuss our findings and offer insight for further research.

\section{Background}

\subsection{Integrated Public Alert Warning System}
\begin{figure*}
	\centering
	\includegraphics[scale=0.5]{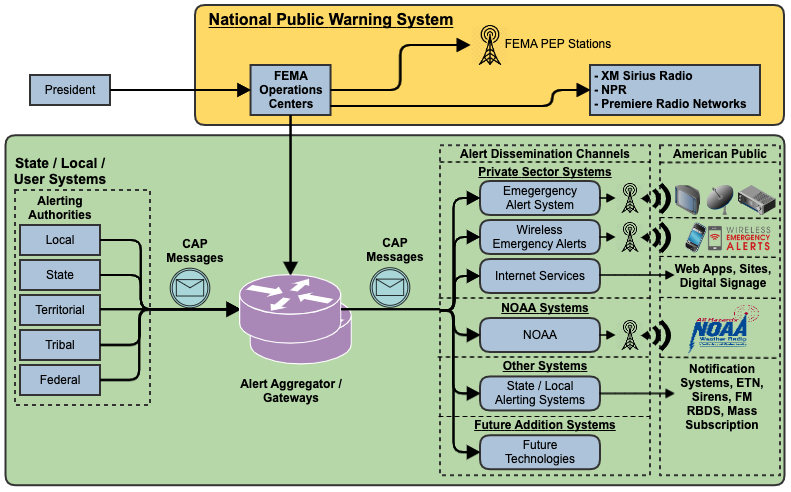}
  \caption{IPAWS-OPEN architecture. }
  \Description{IPAWS-OPEN architecture.}
	\label{fig:ipaws}
\end{figure*}
In compliance with the FCC, the current network designed by FEMA is the Integrated Public Alert Warning System Open Platform for Emergencies (IPAWS-OPEN), which designates how emergency alerts are sent and received by the designated party. IPAWS-OPEN's hierarchy is shown in Figure~\ref{fig:ipaws} with the alerting parties to the left and the recipients on the right. This architecture was constructed by FEMA~\cite{ipaswsopen} to structurally organize and standardize how the national public can be appropriately alerted in times of emergencies. Common Alerting Protocol (CAP) messages from the alerting authorities all pass through the IPAWS-OPEN router, which will distribute the CAP messages to the appropriate alert dissemination channels. Our paper will specifically focus on the Wireless Emergency Alerts block located under the Private Sector Systems dissemination channel. 

\subsection{History of WEA Messages}
\begin{figure}[!htb]
	\centering
	\includegraphics[scale=0.22]{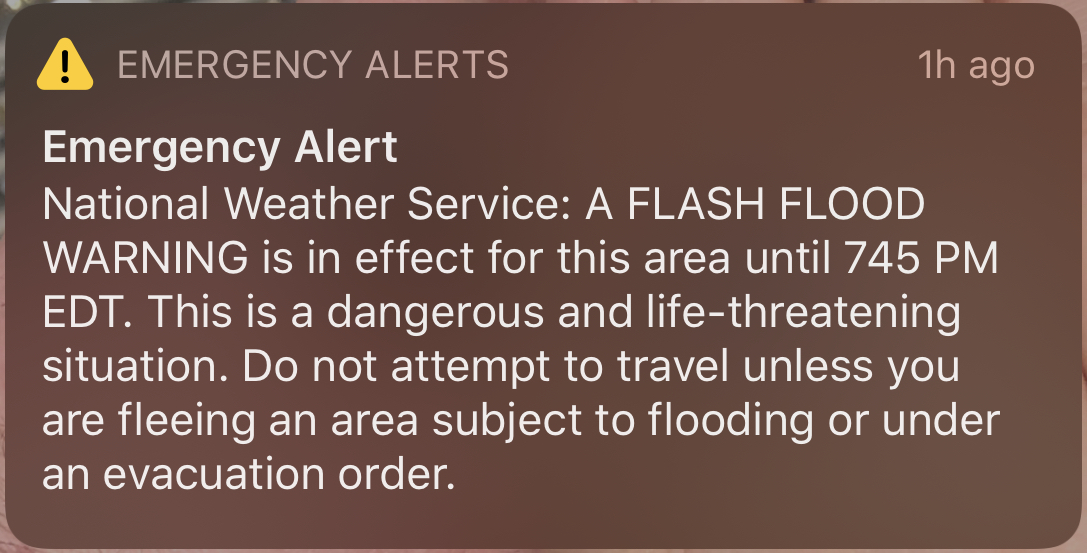}
  \caption{WEA message received for Flash Flood alert July 2020.}
  \label{fig:wea}
  \Description{WEA message received for Flash Flood alert July 2020.}
\end{figure}
WEA messages were initially introduced under the Warning, Alert, and Response Network (WARN) Act of 2006~\cite{moore2008emergency} and with WEA 1.0 became functionally available after 2012~\cite{Wireless28:online}. WEA messages were initially limited to 90 characters complying with the CAP messages at the time and were absent of graphic elements in messages. WEA 2.0 upgrades were announced in 2016, proposing increasing the message length to 360 characters, allowing transmission of embedded links or multimedia, and supporting Spanish translated alert messages~\cite{fcc127:online}. Cell sector geo-targeting of WEAs and the addition of embedded links were introduced in 2017~\cite{WE23:online}. Proposed Spanish functionality officially released in November 2018, with the WEA 2.0 being fully functional by May 2019. WEA 3.0 was outlined in early 2018, proposing changes to improve alert delivery accuracy by narrowing geo-targeting requirements and providing alert preservation for 24 hours~\cite{fcc18:online}. In 2018, alerting authorities sent $7098$ WEA messages~\cite{CTIAIndu63:online} across the United States alerting about emergency weather events. 

\subsection{First Responder Networks}
Following the events of 9/11, FEMA went with AT\&T~\cite{HistoryF67:online} to initialize a network for first responders to be able to communicate over LTE on Band 14 of the $700\ MHz$ spectrum~\cite{FirstNet55:online}. AT\&T titled their service as FirstNet, which should, in critical scenarios, allow first responders to communicate through a dedicated channel with increased priority over civilian traffic. However, even with a dedicated network, first responders have utilized the help of civilians using their cell phones without FirstNet capabilities to assist in rescue efforts~\cite{wax2017cajun}. This is under the assumption that all cellular towers are not damaged or destroyed due to weather conditions, and that the remaining towers are not overloaded with increased traffic due to civilians performing bandwidth-intensive tasks such as uploading videos of weather events or streaming media coverage over their cellular data. 

\section{Related Work}

The design of WEAs is critical to their effectiveness, with the recipients appropriately following received instructions. Previous work has shown that if enough information is not provided during emergencies, people will seek out desired knowledge of the emergency, potentially causing delays in public response~\cite{hardin2017risk, kuligowski2013national}. WEA messages are terse messages by design, meaning that the customization that alerting authorities can implement is limited. When customizing messages, alerting authorities must take caution not to degrade the accuracy or understandability of the alerts or warnings sent out to people in the emergency location. Previous research has examined the best practices for crisis alerts on social media platforms with terse messaging~\cite{bean2016disaster, brynielsson2018informing}. 

\subsection{Response to Alert Messages } 
Previous research has indicated that the social and cognitive responses to warnings and alert messages can be modeled as a multistage process~\cite{lindell2012protective, mileti1990communication, wogalter1999organizing} with the base of this research drawing from theories of collective behavior~\cite{blumer1951collective}, emergent norms~\cite{turner1957collective}, and reasoned action~\cite{fishbein1977belief}. Upon receiving an alert message, there are six characteristics of a person's response to the received message to determine its effectiveness: 
\begin{enumerate}%%[leftmargin=.3in]
	\item Reading the alert
	\item Understanding the risk and severity of an alert provided~\cite{casteel2016assessing}
	\item Believing the alert to be credible~\cite{silic2013information, bean2016disaster, liu2017picture}
	\item Personalizing the emergency with themselves or others around them~\cite{bean2016disaster, liu2017picture}
	\item Deciding to take protective action~\cite{casteel2016assessing, liu2017picture, wood2018milling}
	\item Confirming the received alert~\cite{lindell2003communicating, mileti1995factors, mileti2000social, mileti1990communication, sorensen2000hazard}
\end{enumerate}  
Messages which are both accurate and requiring low cognitive demand to understand have a higher probability of recipients taking appropriate action or response~\cite{sutton2014warning}.

Trust is an undeniably important factor in behavioral response, as a lack thereof may lead people to ignore the contents and risks associated with an alert or warning~\cite{drabek1969social, drabek1971disaster, fritz1954norc, vermeulen2014understanding, stoddard2014wireless, woody2014maximizing}. If the trust between a person and the alerting authority is diminished due to repeated false-alarms or non-factual information, people may exhibit the 'cry-wolf' phenomenon by becoming non-compliant to all future alerts due to believing them to be false~\cite{dow1998crying, sorensen2007community,bradford2017there}. If people feel that they no longer trust alerts, they have the ability on newer devices to disable alert notifications from amber alerts, extreme threats, and severe threats.

Avoidance of frequently repeated alerts when possible must be taken as long-term exposure to repeated messages may cause the recipient to become habituated or suffer message fatigue resulting in the inability to grasp and illicit an appropriate response to a message or disregarding the contents altogether~\cite{thorley2001habituation, wogalter2006comi}. The cognitive phenomenon of habituation is where repeated stimuli become increasingly ineffective at eliciting recipients' attention~\cite{mayhorn2014warning, wogalter2005providing}. Recipients who have become habituated to an alert may ignore the alerts without comprehending or reading the content~\cite{bohme2011security, egilman2006brief, akhawe2013alice, mccroskey1971introduction, stephens2013organizational, bradford2017there}.

\subsection*{Design of Alerts}   
The design of effective alerts needs to ease decision-making~\cite{felt2012android}. If the WEA contents are too complicated, requiring too much cognitive demand, the desired behavior might often be lost. The addition of visual stimuli to alerts may improve the actionability of received alerts~\cite{bean2014comprehensive}. Color has been shown to effectively affect our cognitive system~\cite{elliot2007color} and even our emotions~\cite{naz2004relationship}. The color red also has been shown to be linked to performance attainment~\cite{elliot2007color}. The addition of sensory inputs such as vibration strength and volume level to emergency alerts can assist sensory impaired civilians to recognize and take action on the alert~\cite{mitchell2010human}. The alerts used in our survey followed the guidelines~\cite{mcgregor2014best} for best practices in designing WEAs however, due to limitations of the current WEA protocol, no features revolving around color have been added to the text of the alerts delivered to participants. 
 
For our study, the alerts' design did not account for users becoming habituated to an alert as they were only presented a single WEA. However, an alert design that would best effectively elicit a positive response through a single alert is optimal as repeated alerts may cause habituation. To avoid message fatigue, the contents of the WEAs we designed for the study were to be easily understandable and structured so that the participant knows what is going on, what is expected of them, and based on the alert they were assigned to what would happen if they comply.

Geo-targeting as a design feature for WEAs is the act of sending emergency alerts to only people in affected areas with higher granularity than city-wide alerting. Targeted alerts have been shown to be effective in mitigating exposure to contaminated water areas~\cite{strickling2020simulation}, and previous research has shown that geo-targeting of alerts can be upgraded utilizing a smartphone's location history~\cite{kumar2018rethinking}. Without using geo-targeting of emergency alerts and with WEAs current limitations, civilians in unaffected areas may end up disabling the alerting service on their smartphones~\cite{kumar2018rethinking}, and frequent exposure may damage the trust and image civilians hold of the alerting authority~\cite{bradford2017there}.   

\subsection*{Use of Social Media Disasters} 

% Seek infomation
Civilians naturally seek out sources of confirmation when unexpected information is received or events presented to them~\cite{auf1989disaster}. In the design of Australia's Public Emergency Warning System, it was noted that if civilians did not receive confirmation, there would be a possibility of them dismissing the emergency alert~\cite{mcginley2006design}. Previous works have examined this phenomenon where social media or microblogging services emerged as destinations to collect information on natural or humanmade disasters such as: 2001 World Trade Center Attacks~\cite{schneider2002web}, 2004 Indian Ocean Tsunami~\cite{liu2008search}, 2005 Hurricane Katrina~\cite{palen2007citizen,torrey2007connected,shklovski2008use}, 2007 Virginia Tech shootings~\cite{palen2009crisis,vieweg2008collective}, 2007 California Wildfires~\cite{shklovski2008finding,sutton2008backchannels}, 2008 Sichuan earthquakes in China~\cite{qu2009online}, and the 2008 Nothern Illinois shootings~\cite{palen2008emergent}. 

Methods of communications are vulnerable to their surrounding infrastructure and electric grids~\cite{simon2015socializing,palen2007citizen}. Amid disasters, civilians attempt to use their phones to share their status with their friends and family~\cite{sutton2008backchannels}. During the 2017 Hurricane Maria in Puerto Rico, civilians traveled to less effected areas for cellular service~\cite{becker2017trying}, and others tried switching telecom providers to those whose infrastructure was less affected by the hurricane~\cite{brown2018puerto}. 

% Altruism in emergencies
Before and after the internet era, civilians have spontaneously volunteered to assist in disasters~\cite{dynes1970organized,fritz1957convergence,kendra2003reconsidering,tierney2002facing}. People will form altruistic communities during disasters to assist with those effected~\cite{fischer1998response,tierney2002facing}. 

% Crisis informatics
The use of information and technology in rescue and recovery operations during disasters is known as crisis informatics, a term coined in 2007~\cite{hagar2007information}. During disasters, social media has been a medium to exchange relevant information to the context of their situation~\cite{sutton2008backchannels,vieweg2010microblogging,bird2012flooding,simon2015socializing}. As such a medium, first responders can utilize social media to crowdsource specific tasks to assist rescue and recovery operations~\cite{dittus2017mass,ludwig2017situated}. Communication between first responders and civilians through social media has shown promising results~\cite{reuter2017towards}. Volunteer civilians can also utilize social media for coordination among other volunteers~\cite{starbird2011voluntweeters,reuter2013combining,white2014digital,kaufhold2016self}. With the amount of information being provided by social media sources, first responders have the ability to detect new issues arising as a result of an ongoing disaster~\cite{sakaki2010earthquake,pohl2015social}. 

The use of social media during disasters has many benefits; however, it is not without flaws. Social media sources lack credibility for their information~\cite{hughes2014social}. Reliable access to the information source and the source not becoming overloaded with information cannot be guaranteed to first responders~\cite{mendoza2010twitter}. Methods to mitigate information overload during disasters has been examined and shown successful by applying filters~\cite{kaufhold2020mitigating}. Organizations behind popular social media lack resources and guidance for first responders to, optimally, utilize the information source~\cite{plotnick2016barriers}.

To the best of our knowledge at the time of writing this paper, the only prior research on the reduction of mobile usage before and after receiving a WEA alert was presented as a poster~\cite{youssef2016message} and demonstrated through a survey that participants might not be willing to reduce their mobile usage. In this paper, we design several alternative framings for WEAs to increase user compliance to message contents. The seven alerts designed were contained basic information similar to current WEAs, altruism to compare with previous research, multimedia, negative feedback, positive feedback, reward, and punishment. 

Our contributions are: (1) we use a simple categorization of 12 most used mobile apps based on their data rate, in order to capture/measure users daily traffic, (2) we also show that the effectiveness of WEAs is varying depending on the wording and objective, for example, basic information provides the largest reduction in hours using multimedia and video service tasks, using punishment in alerts can diminish compliance, positive reinforcement has the most considerable absolute compliance to alerts.

\section{Survey of US WEA response}

\subsection{Recruitment}
We recruited a total of 898 participants through Amazon's Mechanical Turk (MTurk) to take part in a survey observing their mobile usage behavior before and after receiving a WEA message in a hypothetical critical weather disaster. Forty-four participants were removed from the analysis due to failing the attention check or choosing to skip required responses, leaving a final sample size of $854$. Participants' recruitment criteria were to be fluent English speakers, smartphone owners, and residents of the United States of America. The duration of the study was determined to be approximately $20$ minutes from the previously launched Pilot study, so to appropriately match the duration with the nation's minimum wage~\cite{Question60:online} participants were awarded $\$2.50$ for completing the survey.
\begin{table*}[ht]
	\centering  
	\includegraphics[scale=0.4]{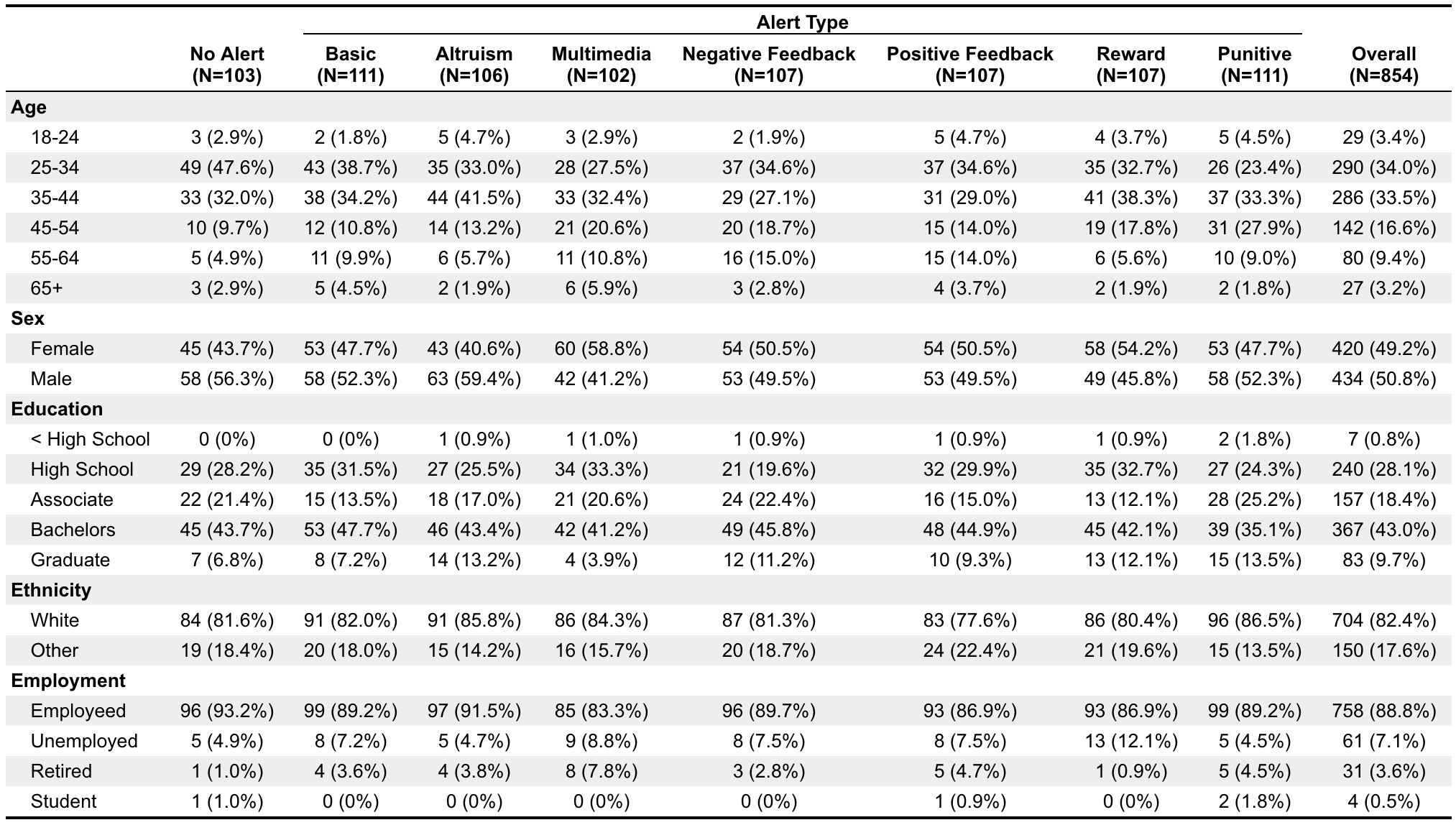}
	\caption{Participant Demographics}
	\label{test}
\end{table*} 
Men comprised $51\%$ of the sample collected, and women comprised the remaining $49\%$ of the sample. Ages ranged from 18 to 65+, with a majority of the participants in the ranges $25-34 (33.96\%)$ and $35-44 (33.49\%)$. The education background of participants was diversely distributed between Less than High School $(0.82\%)$, High School $(28.10\%)$, Associates Degree $(18.38\%)$, Bachelors Degree $(42.97\%)$, and Graduate School $(9.72\%)$. 

A between-subjects design was made, dividing participants into one of 7 groups or a control group where no alert is received at all. The seven alert types were Basic Information, Altruism, Multimedia, Negative Feedback, Positive Feedback, Reward, and Punishment. The message displayed to the participants can be seen in Figure~\ref{fig:alerts}; each participant received only one of the alerts unless they were randomly assigned to the control group. Those in the control group received no alert message and were displayed the following "During natural disasters like hurricanes, after observing the change in environment, how frequently do you think you would perform each of these tasks?"  Of the remaining $854$ participants, $104$ were assigned to the Altruism alert, $102$ were in the Multimedia alert, $111$ were in the Basic alert, $107$ were in the Negative Feedback alert, $107$ in the Positive Feedback alert, $111$ were in the Punishment alert, $107$ were in the Reward alert, and $103$ were in the control group receiving no alert during the study.

\subsection{Procedure}  
\begin{figure}[htb]
	\centering
	\includegraphics[scale=0.5]{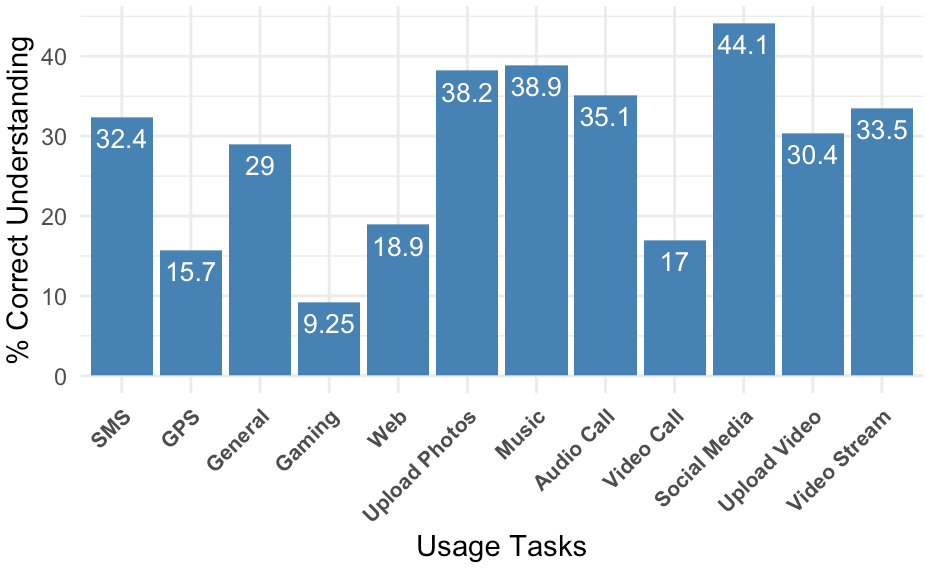}
	\caption{Percentage of Correct understanding of data consumption.}
  \label{fig:understand}
  \Description{Percentage of Correct Understanding}
\end{figure}

First, participants completed an informed consent form approved by our University's Institutional Review Board. In the next stage in the study, participants reported the amount of time they best estimated to perform each of the 12 bandwidth-intensive tasks on an average day and were tasked with categorizing tasks according to how much network resources are consumed in terms of their descriptive labels "Very Low (1-10 texts)" to "Very High (10,001+ texts)". The number of text messages listed next to each of these options is there to help participants conceptualize what the descriptive label means in terms of data, as a well-known reference, considering that some technology illiterate participants may not be familiar with the notion of bandwidth as demonstrated in Figure~\ref{fig:understand}. All participants received this stage before being randomly assigned to either the control group or one of the seven groups, which represent possible improvements that can be applied to the WEA branch of IPAWS-OPEN.

After being randomly assigned to their groups, participants are then being asked, for each of the 12 bandwidth-intensive tasks, whether they would now use the task "More than before the alert," "Same as before the alert," "Less than before the alert," or  "Never (I won't perform this task at all)." To test the reliability of responses and to make sure that participants were appropriately answering this to their utmost ability, we required users to also respond with how long they estimated they would utilize the 12 tasks in a similar fashion to the first part of the survey, and paired an attention check in the middle of the survey. The participants' response on how long they would perform tasks after receiving alerts (More than before, Same as before, Less than before, or Never) helps us to calculate the total time duration of users usage of each application, which then is interpreted as the estimated data of users from using each application. 

We categorized participants' compliance to the alert using three levels for each task, which includes partial compliance ("Less than before the alert" is selected), full compliance ("Never (I won't perform this task at all)" is selected), and no compliance ("Same as before the alert" or "More than before the alert" is selected). Also, we characterize the situation in which the participants show full compliance for all the tasks as the absolute compliance level. 

\subsection{Bandwidth Consumption Tasks}
\begin{figure*}
	\centering
	\includegraphics[scale=0.45]{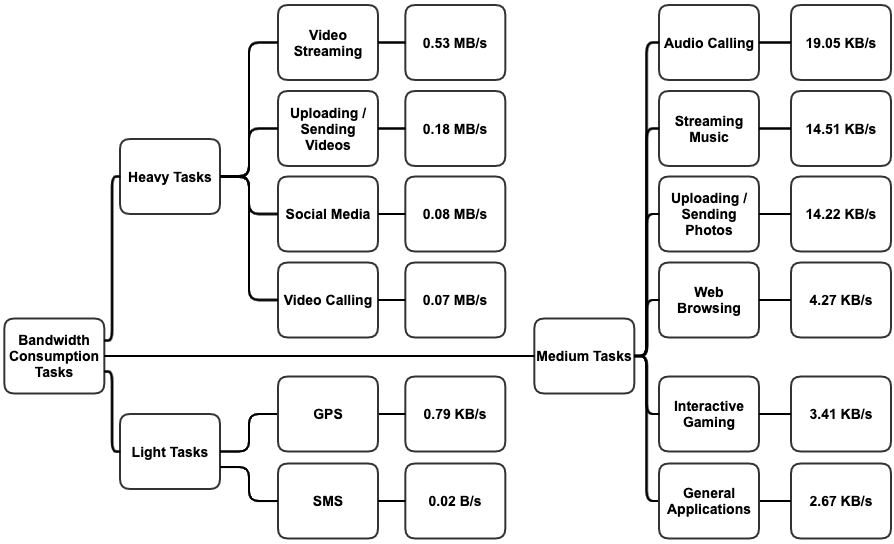}
	\caption{The bandwidth intensive tasks which can be performed on a smartphone are organized to categories of Heavy Tasks, Medium Tasks, Light Tasks. The average data rate per second is provided next to each task and adapted from usage data calculators from AT\&T~\protect\cite{ATTDataC29:online}, Sprint~\protect\cite{SprintDa19:online}, and Verizon~\protect\cite{Househol39:online} as they make up a majority of cellular subscriptions in the US~\protect\cite{USwirel42:online}.}
  \label{fig:dataconsumption}
  \Description{Bandwidth consumption tasks organized into Light, Medium, and Heavy tasks. Data rates provided based on usage calculators from AT\&T, Sprint, and Verizon.}
\end{figure*}
As shown in Figure~\ref{fig:dataconsumption}, the tasks which smartphone users perform can divide into three categories: heavy tasks, medium tasks, and light tasks. Heavy tasks involve those tasks which are video or multimedia-related, consuming the most network resources, and quickly consuming multiple MegaBytes (MBs) of data within a minute of use. Heavy tasks also include social media such as Facebook, Instagram, and Twitter, as they are frequently saturated with video content. Medium tasks, such as uploading photos or streaming music, are a large part of daily smartphone use consuming a magnitude of data less than those in the Heavy task category.  Light tasks are those tasks that consume such a small amount of network resources from cellular towers that having $100\%$ of users connected to a tower perform the task should not overload the tower capacity. The bandwidth consumed for each of the tasks were extracted from averages provided by AT\&T~\cite{ATTDataC29:online}, Sprint~\cite{SprintDa19:online}, and Verizon~\cite{Househol39:online}. The 12 bandwidth-intensive tasks were created by aggregating the different service providers' application categories to build an inclusive list of all types of mobile applications; however, the novelty is more in using these 12 categories in estimating the total generated cellular traffic of participants.

\subsection{Alert Types}
\begin{figure*}[ht]
	\centering 
	\subfigure[Basic Alert]{
		\includegraphics[scale = 0.5]{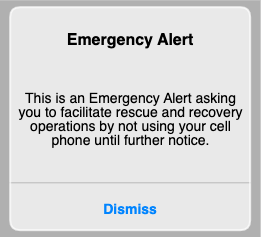}
		\label{fig:information}
	}
	\hfill
	\subfigure[Altruism Alert]{
		\includegraphics[scale = 0.5]{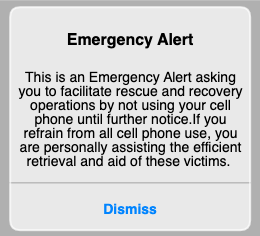}
		\label{fig:altruism}
	}
	\hfill
	\subfigure[Multimedia Alert]{
		\includegraphics[scale = 0.5]{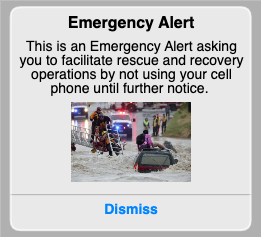}
		\label{fig:multimedia}
	}
	\hfill
	\subfigure[Negative Feedback Alert]{
		\includegraphics[scale = 0.5]{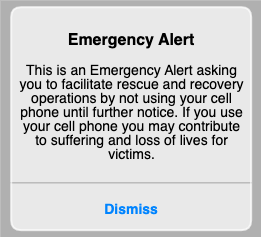}
		\label{fig:negativereinforcement}
	}
	\hfill
	\subfigure[Positive Feedback Alert]{
		\includegraphics[scale = 0.5]{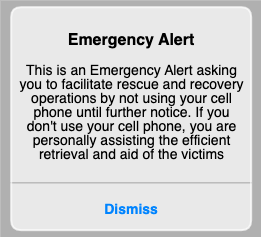}
		\label{fig:positivereinforcement}
	}
	\hfill
	\subfigure[Reward Alert]{
		\includegraphics[scale = 0.5]{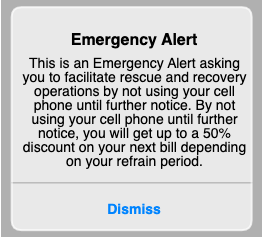}
		\label{fig:reward}
	}
	\hfill
 
	\subfigure[Punishment Alert]{
		\includegraphics[scale = 0.5]{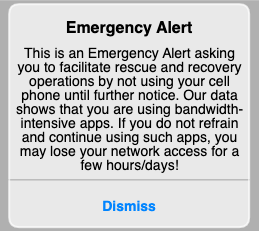}
		\label{fig:punitive}
	}
	\hfill
	\caption{Alerts received by individuals during the survey. Participants were assigned to one of the alerts or a control condition.}
  \label{fig:alerts}
  \Description{Alerts received by participants}
\end{figure*}

The wording of alerts serves the objective for each alert type, for example, altruism. The list of alert types in this survey, as well as their respective wording, is derived from consulting with a team of professors from our University's Psychology Department.

\textbf{Basic Information Alert.} As seen in Figure~\ref{fig:information}, basic information is provided to participants stating just that there is an emergency and that the participant should refrain from using their device. This alert is similar to alerts currently received when weather conditions are alerted with an added message of reducing their network usage. 

\textbf{Altruism Alert.} The Altruistic alert that participants could be assigned to appeals to users that they are personally going to benefit the aid of victims of emergencies by restraining from using their mobile devices. Altruistic messages have previously been noted as an effective method for reducing usage during times of emergencies~\cite{youssef2016message}.

\textbf{Multimedia Alert.} The Multimedia alert is utilizing the newest addition to WEAs that they have the capability of adding images to weather alerts. To determine the effectiveness of this addition, we designed an alert that provides basic information while requesting that the user restrains from utilizing their mobile device and providing an image of a weather disaster.

\textbf{Negative Feedback Alert.} The alert attempts to utilize Negative Feedback by having the user reduce their mobile usage in response to avoiding the adverse stimuli of potentially contributing to the suffering of victims in an emergency. The term Negative Feedback comes from the work of B. F. Skinner in 1963~\cite{skinner1963operant}.
 
\textbf{Positive Feedback Alert.} The Positive Feedback alert provides information that the participant should refrain from cell phone usage, and they will know that they are helping out other civilians as positive stimuli. The wording is similar to that of the Altruism alert; however, it was carefully chosen to demonstrate the effects that wording has on WEA compliance.
	
\textbf{Reward Alert.} The Reward alert tries to incite the participant with financial benefits for complying with the WEA during an emergency. Participants are incentivized with a discount off their next phone bill, dependent on how much they comply with the alert. 

\textbf{Punishment Alert.} The Punishment alert added a punishment message to the basic information. Users were told that they needed to restrict their device usage until further notice, and their punishment for denying this is that they would lose access to their service. This punishment is to incentivize fear in order to obtain compliance. 

\section{Experiment: Results and Discussion}

\subsection{Participants Appear to Represent the General Population} 
Participants were $49.2\%$ and $50.8\%$ Female and Male, respectively, comparing with U.S. Census Data~\cite{USCensus69:online}, with the division being around $50.8\%$ and $49.2\%$ Female and Male, respectively. The U.S. population's ethnicity is about $76.3\%$ White and $23.7\%$ other~\cite{USCensus69:online}; similarly, our sample population is $82.4\%$ White and $17.6\%$ other. Although in terms of age~\cite{AgeandSe66:online} and education~\cite{Educatio2:online} of the participants, we have observed some divergence from the general U.S. population, which statistically could be a result of having a smaller sample space.

The average data predicted to be used per month per smartphone in the US through 3G/4G/LTE is around $13$ GB in 2019, equating to around $0.45$ GB/day, not including mobile data that is passed through WiFi networks~\cite{Ericsson74:online}. Using the estimated data rates in Figure~\ref{fig:dataconsumption}, the sum of all participants' predicted usage per month is around $70.4$ GB, equating to $2.34$ GB/day. To understand the divide of mobile traffic between how much traffic is passed through cellular networks or WiFi, we find from~\cite{CiscoAnn95:online} that an average user will consume around $25\%$ of their mobile traffic through cellular and the other $75\%$ through WiFi. Applying this ruling to participant data, the average mobile data consumed on the cellular network per month is around $17.5$ GB/day, equating to around $0.58$ GB/day. This rate is assuming that not all videos are viewed over a WiFi network and that the participants continue to use, on average, $8.5$ hours per device per day for the entire month.

\subsection{On Understanding of Usage Tasks}
To better understand participants actions, we observe their understanding of how much impact the usage of the 12 mobile application tasks have on a network comparing with basic text messages. Participants were asked for each task: "On average, how do you think this task consumes network resources (bandwidth) per second as compared to sending text messages? (max length text with 160 characters)." They were then instructed to choose from the options: "Very Low (1-10 texts)," "Low (11-100 texts)," "Average (101-1000 texts)," "High (1001-10,000 texts)," and "Very High (10,001+ texts)."

Among all the usage tasks in question, people had the best understanding ($44.1\%$) of how much data is consumed by Social Media per second of use. We further note that for all usage tasks, less than half of participants were correctly able to choose the correct average data consumption. These results provides us relevant information about: (1) the participant's understanding about how much data is consumed for particular tasks as such information could impact their level of compliance as it demonstrated that participant's knowledge for bandwidth-consumption of different tasks is not accurate, and (2) to better increase the compliance of users it might be useful if we use well known references like text messaging to describe applications data rate.

\subsection{On Compliance to WEAs}
\begin{figure}[htb]
	\centering
	\includegraphics[scale=0.5]{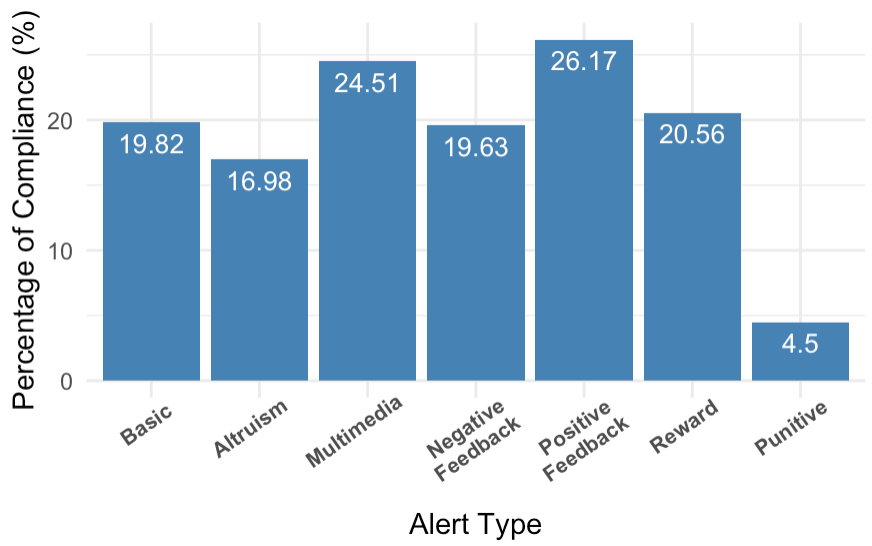}
	\caption{Absolute Compliance by Alert Type}
  \label{fig:abscomp}
  \Description{Absolute Complian by Alert Type}
\end{figure} 
\begin{figure}
  \centering 
  \subfigure[Partial Compliance by Usage Task]{
    \includegraphics[scale=0.43]{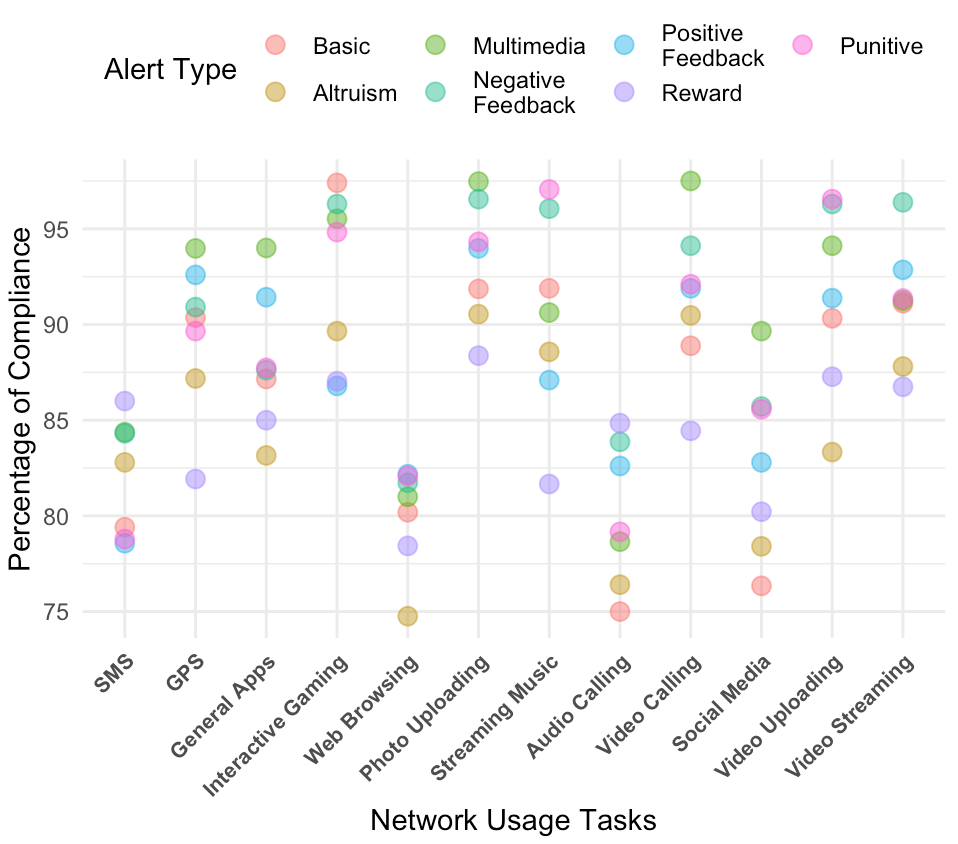}
    \label{fig:partial}
  }
  \subfigure[Full Compliance by Usage Task]{
    \centering
    \includegraphics[scale=0.43]{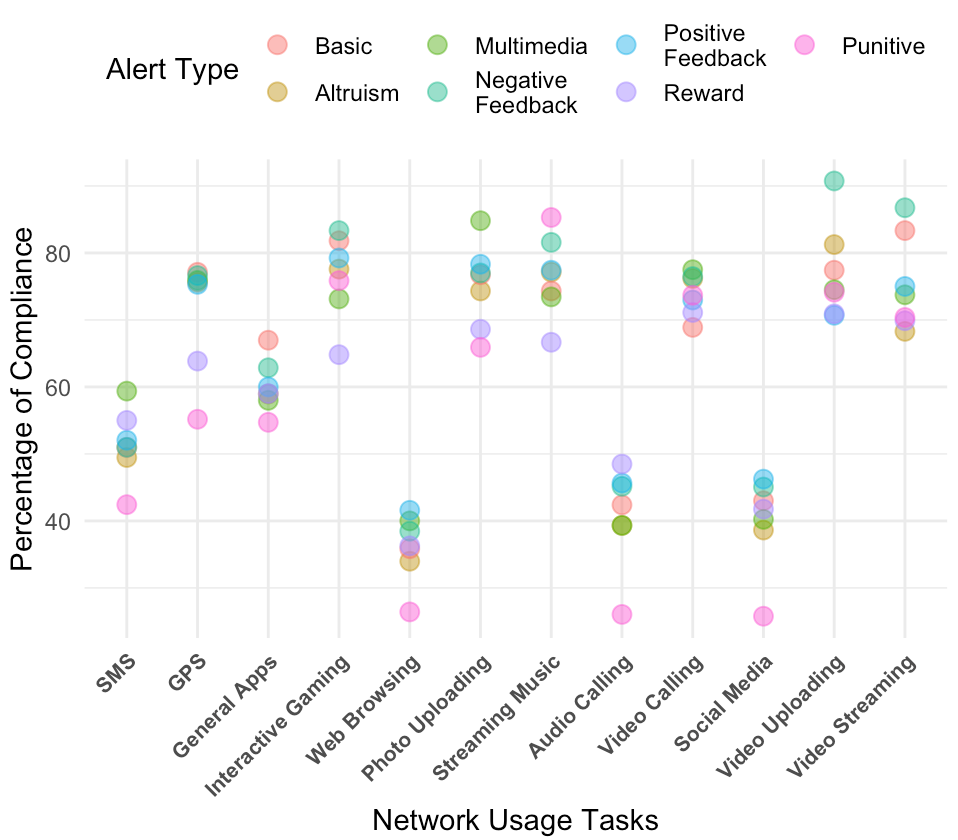}
    \label{fig:full}
  }
  \caption{Full and Partial Compliance to received alerts}
  \Description{Full and Partial Compliance to received alerts}
\end{figure}
Compliance to alerts was explored as partial compliance, full compliance, and absolute full compliance. Partial compliance, as seen in Figure~\ref{fig:partial}, was defined for participants that would reduce their usage of particular tasks once they saw their alert. Full compliance, as seen in Figure~\ref{fig:full}, was defined as users who would completely restrict all usage for particular tasks. Absolute compliance, as seen in Figure~\ref{fig:abscomp}, was defined for participants who would stop the usage of all heavy, medium, and light tasks until they are further updated on the situation.

Participants who were assigned Negative Feedback and Basic Information alerts demonstrated the highest percentage of compliance to their alerts, as can be seen in Figures~\ref{fig:partial} and \ref{fig:full}. In regards to the heaviest task Video Streaming, $91.11\%$ of information participants claimed they would at least partially comply with restrictions, and $83.33\%$ would fully comply with fully restricting their usage. Participants showed the least compliance for SMS, Web Browsing, Audio Calling, and Social Media usage after receiving an alert which follows prior research that civilians will increase usage of information finding tasks after disasters~\cite{palen2009crisis,vieweg2008collective,qu2009online,palen2008emergent,shklovski2008finding,sutton2008backchannels,palen2007citizen,torrey2007connected,shklovski2008use,liu2008search}.

\subsection{Participants Who Usually Read Alerts Reduce Their Usage More}
A one-way multivariate ANOVA was conducted to compare the effect of the independent variable (IV) participants self-reporting that they read WEAs received on their smartphone on the dependent variables (DVs) their time reduced using heavy, medium, and light bandwidth-intensive tasks during weather disasters. There was a statistically significant effect of reading alerts on the reduction of bandwidth-intensive tasks, $F(3,850)=4.4687$, $p<0.005$; Wilk's $\Lambda = 0.9845$, partial $\eta^2 = 0.0155$. The observed response behavior is consistent with the six characteristics~\cite{sorensen2000hazard} of alert effectiveness; hence we should design alert systems to deliver WEAs in a way that best catches their attention.

\begin{figure*}[ht]
	\centering 
	\includegraphics[scale=0.44]{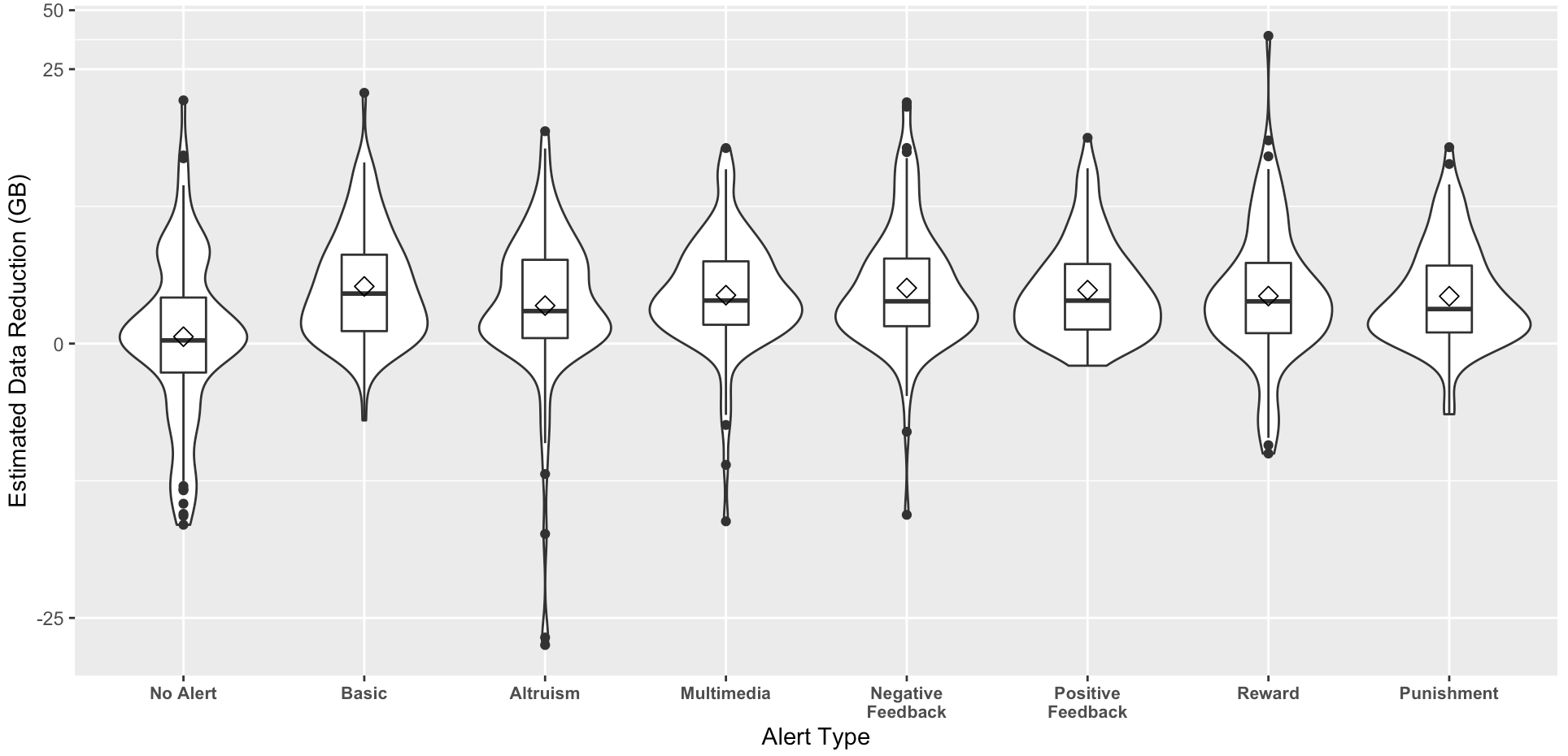}
  \caption{Estimated reduction of data consumed in GigaBytes (GBs) by Alert Type}
  \label{fig:reducthours} 
  \Description{Estimated reduction of data consumed in GigaBytes (GBs) by Alert Type}
\end{figure*} 
\subsection{Participants With No Alert Utilize High-Bandwidth Tasks During Disasters}
Participants who were assigned to any alert type, on average, reported that they would reduce their Heavy task consumption by 91.68 minutes after receiving a WEA on their smartphone. This is in contrast to those participants who received no alert at all but informed of bad weather conditions outside reported that they would increase their usage of Heavy tasks by an average of 15.83 minutes. These findings highlight the results observed in Figures~\ref{fig:reducthours},~\ref{fig:medhours}, and~\ref{fig:heavhours}, which shows that using any WEA regardless of the type could decrease the users generated cellular traffic as opposed to no alert; hence any alert is better than no alert. 

\subsection{On the Effect of Alerts on Heavy, Medium, and Light Tasks}

Light tasks (GPS, SMS), Medium tasks (Gaming, Upload Photo, General, Audio Call, Web Use, Music), and Heavy Tasks (Social Media, Video Call, Upload Video, Video Stream) were grouped to see if there any significant effect from any of the alerts and three task categories. Three omnibus ANOVA tests were conducted on the change in minutes of the three task categories to determine if there would be interest in delving deeper into alerts' effects on the overall categories. It was noted that there was no significant difference between alerts groups in the Light task category $F(7,846)=1.4181$, $p>0.05$. Medium and Heavy tasks were found to have significant difference between alerts $F(7,846)=6.7835$, $p<0.000$ and $F(7,846)=5.8207$, $p<0.000$, respectively. 

\begin{figure}
  \includegraphics[scale=0.55]{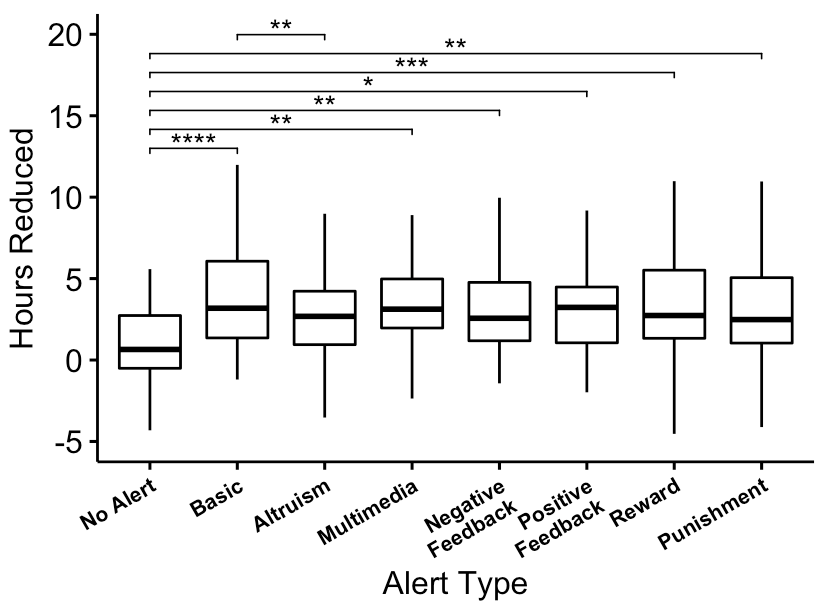}
  \caption{Hours Reduced per Alert Type for Medium Tasks}
  \label{fig:medhours}
  \Description{Hours Reduced per Alert Type for Medium Tasks}
\end{figure}
To analyze the difference between alerts for both Medium and Heavy tasks pairwise t-tests with Bonferroni correction were implemented. For Medium Tasks, between being assigned an alert or not, being assigned an alert containing just Basic Information showed the largest significant reduction of minutes of utilizing tasks; $t=239.2933, p<0.000$. This is followed by receiving a Reward alert $t=180.6241, p<0.000$, a Multimedia alert $t=175.1844, p<0.000$, a Negative Feedback alert $t=161.1661, p<0.00$, a Punishment alert $t=159.0320, p<0.00$, a Positive Feedback alert $t=158.2876, p<0.00$. Receiving an Altruistic alert showed no significant difference in the reduction of minutes of utilizing Medium Tasks from receiving no alert; $t=91.2696, p>0.05$. Receiving just an alert containing information showed a significance in the reduction of network consumption during an emergency than the Altruistic alert with $t=148.0236, p<0.00$. The results of these pairwise t-tests can be graphically observed in Figure~\ref{fig:medhours}.

\begin{figure}
  \includegraphics[scale=0.55]{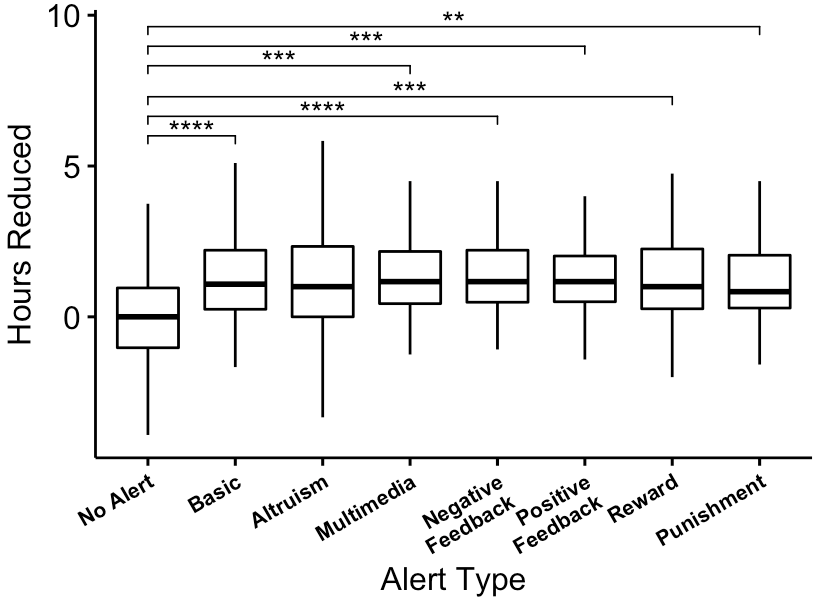}
  \caption{Hours Reduced per Alert Type for Heavy Tasks}
  \Description{Hours Reduced per Alert Type for Heavy Tasks}
  \label{fig:heavhours}
\end{figure}
For Heavy Tasks, between being assigned any alert and not, receiving an alert containing Basic Information showed the largest significant reduction in minutes of utilizing bandwidth-intensive tasks during a hypothetical emergency; $t=128.6277, p<0.000$. This was followed up by receiving a Negative Feedback alert $t=122.0686, p<0.000$, a Reward alert $t=116.0406, p<0.000$, a Positive Feedback alert $t=115.1153, p<0.000$, a Multimedia alert $t=111.8644, p<0.000$, and a Punishment alert $t=102.5647, p<0.00$. As with the Medium Tasks, receiving an Altruistic alert has no significant difference from receiving no alert at all; $t=55.4104, p>0.05$. These findings consistently hold to altruism's possible exclusion from possible enhancements to the wording of WEAs. Results can graphically be seen in Figure~\ref{fig:heavhours}.

\section{Conclusion}
In conclusion, we have used 12 categories of mostly used cell phone application types to estimate users' daily traffic and to see the impact of alerts on reducing this. It has been observed that providing any alert to users is more helpful with reducing the load on the cellular network than no alerts, as it's been shown in the results. Also, the different wordings of alerts and their objectives could have an impact on their effectiveness in reducing non-essential cellular traffic. We noticed that just providing basic information regarding the disaster can significantly reduce cellular network consumption for both Medium and Heavy Tasks, which are the main contributers to cellular traffic; also, we have seen that users who usually read alerts showed higher compliance levels. Appealing to Altruism in WEA alerts showed little at all ability to reduce participants network consumption, which is consistent with prior research on the topic~\cite{youssef2016message}. Two possible explanations for altruism's lack of effectiveness could be the addiction of users to cellphone usage and people's natural habit of seeking news through different cellphone applications during disasters.  

Regarding the relationship between message lengths of WEAs and the recipients' compliance, the most effective alert type was the Positive Feedback alert, which has a higher text length compared to the Basic Information alert. However, larger alerts do not necessarily increase compliance; as we can see, the Punitive alerts have much less compliance compared to Basic Information alerts. The effectiveness of alerts depends on multiple variables, as discussed in the Related Work. The lengths could have an impact on these variables; for example, if the alerts are too long, there is an increased probability of users not reading alerts. 

Current WEA messages that are frequently sent, as seen in Figure~\ref{fig:wea}, does not inform recipients that they should restrict from frivolous smartphone usage during emergencies. The enhancements proposed in this paper showed significant promise at reducing mobile data when compared with participants who received no alert. In addition, the proposed modifications to WEAs in this paper are possible with no extra hardware modification but with the improved design of the message body on the part of alert authorities when constructing CAP messages to send through IPAWS-OPEN.

\subsection{Future Work}
Pros of our crowdsourced approach are that it allows us to gain an initial understanding of possible enhancements to WEAs and potentially reduce bias and increase validity as participants were not interacted with directly~\cite{paolacci2010running}. 

Possible limitations of the survey are that we did not have a way to measure the effect duration of alerts and the durations of usage specified by users are rough estimates of their daily usage. A possible future approach would to be to conduct a field study to actually determine how long compliance will last after receiving an alert and whether this matches the predicted behaviors of participants or in the case of hypothetical situations we should find ways to interpret data in a more realistic fashion and remove potential biases on the survey. This will also provide the ability for us to determine how long after receiving the alert does it take for a user to: (1) read the alert and (2) actually reduce usage after reading. As for the second limitation, this would also be solved through a field study so we would be able to obtain mobile data usage through participant devices accurately and also determine habituation effects of WEAs after repeated exposure.

\begin{acks}
  This work is supported in part by the NSF under Grant No. ACI-1541069. The authors thank Arnold Glass and Margaret Ingate from the Rutgers Psychology department for their insights and comments that improved the design of the survey used in this study.
\end{acks}

\bibliographystyle{ACM-Reference-Format}
\bibliography{sample}

\end{document}